\documentclass[aps,prl,amssymb,twocolumn]{revtex4}

\usepackage{graphicx, color}

\newcommand{\bea}{\begin{eqnarray}}

\newcommand{\eea}{\end{eqnarray}}

\newcommand{\beq}{\begin{equation}}

\newcommand{\eeq}{\end{equation}}

\newlength{\textwidthm}

\begin{document}

\title{Collisional cooling of ultra-cold atom ensembles using 
 Feshbach resonances}

\author{L.~Mathey, Eite Tiesinga, Paul S. Julienne, Charles W. Clark}

\affiliation{Joint Quantum Institute, National Institute of Standards and Technology and University of Maryland, 
 Gaithersburg, Maryland}

\date{\today}

\begin{abstract}
We propose a new type of cooling mechanism for ultra-cold 
 fermionic atom ensembles, 
 which capitalizes on the energy dependence of inelastic collisions 
 in the presence of a Feshbach resonance.
 We first discuss the case of a single magnetic
 resonance, and find that the final temperature and the cooling rate is
 limited by the width of the resonance.
  A concrete example, based on a $p$-wave resonance of $^{40}$K, is given.  
We then improve upon this setup by using both a very sharp optical
 or radio-frequency induced 
 resonance and a very broad magnetic resonance and show that one can
 improve upon temperatures reached with current technologies.
\end{abstract}

\maketitle


The technology of cooling atomic ensembles has been one of the 
 most important developments in physics over the last decades 
 \cite{phillipsnobel}. 
Cooling atomic samples has two ingredients. The first is a ``knife''
that selectively removes atoms with the largest kinetic energy. Secondly,
elastic collisions between the atoms thermalize the remaining atoms.
 It has been a critical ingredient in creating Bose-Einstein 
 condensates \cite{BEC}, in improving atomic clocks \cite{phillips}, 
 and studying atomic properties \cite{lett}. 
The temperatures that have been achieved in bosonic gases are now
 well below a nano Kelvin. For fermionic systems, however, temperatures
 are just below micro Kelvin, due to limitations of current technology.  
 There is therefore interest in developing improved
  cooling methods for fermions.

  We propose a new type of cooling mechanism that uses inelastic
  scattering processes due to a narrow magnetic Feshbach resonance
  \cite{Tiesinga1993,Inouye1998}.
  Here, a molecular state that is resonantly coupled to two scattering
 atoms acts as a ``knife'' that changes the internal state of
  colliding atoms  in a strongly energy-selective
  manner. These states are then either untrapped or have gained sufficient
 kinetic energy to quickly leave the trap. 
 Elastic scattering near the resonance leads to
  thermalization.  For a resonance to generate loss, the
  internal state of the atoms cannot be the lowest energy state, as
  there need to exist states into which they can scatter.  
 For samples of atoms in their lowest
  state,   
 we further propose combining
  a narrow optical \cite{Fedichev1996} or radio-frequency (rf) induced \cite{RFinduced} 
resonance with
  a broad magnetic resonance.  Here, the narrow resonance
  generates the loss processes, whereas the broad magnetic resonance
  drives thermalization.  We will show that we can reach lower
  temperatures than feasible with a single resonance.


%
%
\begin{figure}
\includegraphics[width=7.1cm]{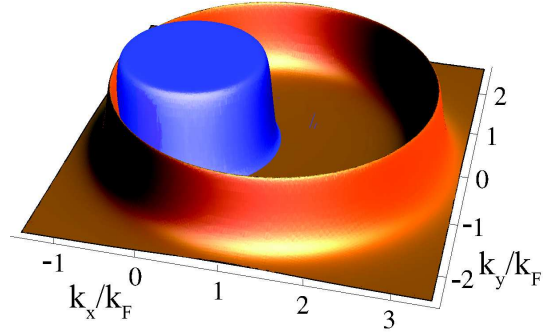}
\caption{\label{FSRes} 
Schematic representation of the loss processes that lead to cooling. 
 We show the momentum distribution $(k_x,k_y)$ 
 of a degenerate Fermi gas (blue, inner
 feature),
  and the inelastic scattering
 rate $K_{in}(\vec{k},\vec{p})$ (orange, ring-shaped feature), both
 in arbitrary units, 
  with a scattering partner with a momentum slightly above the
 Fermi surface at $\vec{p}=(p_x, p_y) = (1.05 k_F, 0)$, and with a resonance
 energy that corresponds to a relative momentum of $1.05 k_F$.
 The losses are largest where the two surfaces approach each other.  
 This corresponds to 
 atoms with momenta located on opposite
 sides of the Fermi sea. 
 (The momenta along $z$ are not shown for clarity.)
}
\end{figure}

Figure \ref{FSRes} shows a schematic representation of the cooling
process. We assume a gas of Fermi atoms in a single state, although
the ideas can be generalized to bosonic or multi-species
and -states fermionic gases.  In the degenerate regime the atoms with
mass $m$ form a Fermi sea, as shown in the figure, with Fermi energy
$E_F = k_F^2/(2m)$, Fermi momentum $k_F$, and a temperature $T$ less
than $E_F/k_B$. We consider a narrow Feshbach resonance, that for two
colliding atoms with momenta $\vec k$ and $\vec p$ induces atom loss
with a rate coefficient \cite{reginaldo94,bohnjulienne99,
  Moerdijk1995, Hutson2007, Napolitano1994}
\bea
K_{in}({\vec k}, {\vec p}) & = & v_r \frac{\pi \hbar^2}{k_r^2}
   \frac{\Gamma(E) \Gamma_0}{(E-E_{res})^2 
 + \Gamma_{tot}^2/4}.
\eea
%
This coefficient is only a function of the relative collision energy
$E$ and is strongly localized around the resonance energy $E_{res}$,
which can be controlled by external magnetic field.  The energy $E$ is
given by $E = \vec k_r^2/(2m_r) = m_r v_r^2/2$, where $\vec k_r =
(\vec k - \vec p)/2$ is the relative momentum and $m_r = m/2$ is the
reduced mass.  Finally, the total energy width is $\Gamma_{tot} =
\Gamma_0 + \Gamma(E)$, where $\Gamma_0/\hbar$ is the linewidth of the
resonant state and $\Gamma(E)/\hbar$ is the collision-energy-dependent
stimulated width.
 
In order to generate cooling, we need to choose the resonance energy
so that only atoms with momenta larger than $k_F$ are lost.  The
largest relative momentum is about $k_F$, corresponding to a pair of
atoms on opposite sides of the Fermi sea.  Consequently, $E_{res}$
must be set slightly above $k_F^2/(2m_r) = 2 E_F$.  The exact amount
that the resonance energy lies above the Fermi energy will be
determined by the temperature of the gas and $\Gamma_{tot}$.
Moreover, the resonance energy needs to be gradually lowered in time
by changing the magnetic field, as the atom number decreases due to
the losses, and thus the Fermi energy decreases.

  At the beginning of the cooling process the temperature is much
  larger than $\Gamma_{tot}/k_B$.  We will find that the smallest
  temperature that can be achieved is $4 \Gamma_{tot}/k_B$.  This
  suggests the use of arbitrarily narrow resonances.  However, the
  thermalization rate is also influenced by the resonance.  In fact,
  the total elastic scattering rate coefficient $K_{el}(\vec k, \vec
  p)$ is
\bea
K_{el}({\vec k}, {\vec p}) & = & v_r \frac{\pi \hbar^2}{k_r^2}
   \frac{\Gamma^2(E)}{(E-E_{res})^2 
 + \Gamma_{tot}^2/4}.
\eea
%
%
%
%
Typically, during the cooling process it is preferrable that the Fermi
gas is close to thermal equilibrium, and therefore we require the
ratio $K_{el}/K_{in} = \Gamma(E)/\Gamma_0 \gg 1$ for $E \approx 2
E_F$.  Consequently we have $\Gamma_{tot}\approx \Gamma(E)$.  Since we
want a temperature that is as low as possible, this leads to competing
requirements on $\Gamma_0$ and $\Gamma(E)$.

For fermionic atoms in the same internal state, only odd partial wave
scattering exists.  In fact, for ultra-cold atoms we only need to
include the $p$ or $\ell=1$ partial wave.  Moreover, the energy
dependence of $\Gamma(E)$ is enforced by the Wigner threshold laws.
Here, this leads to $\Gamma(E) = A E^{3/2}$, where $A$ is an intrinsic
property of the resonance.

In Ref.~\cite{Gaebler}, a $p$-wave resonance  was characterized
 in the collisions of fermionic
 $^{40}$K atoms in the hyperfine state $|9/2,-7/2\rangle$. 
 For the $m_l=0$ component of the $p$-wave resonance,
  they find $\Gamma_0/k_B = 0.9$ nK  
 and $\Gamma(E)/k_B = 1.4 \times 10^{-3}$ nK at $E/k_B = 1$ nK.

   {\it Classical limit.}  Before we consider cooling a Fermi gas in
   the degenerate regime, we treat the classical limit of a
   three-dimensional homogeneous gas, in which we assume
 that  the
   momentum distribution $f(\vec p, t)$ remains
  a Maxwell-Boltzmann distribution
  throughout the
   cooling process\cite{anderlini}.  In fact
\bea
f(\vec p,t) & = & \varpi(n,T) \exp(- p^2/(2 m k_B T)),
\eea
 where  $\varpi(n, T) = (2\pi \hbar)^3 n/(2\pi m k_B T)^{3/2}$ is the phase 
 space density, 
 and only
 the particle density $n$ and the temperature $T$ are time dependent.
 Its time evolution is
\bea
\partial_t f(\vec p,t) 
 & = & 
-  \int \frac{d^3 k}{(2 \pi \hbar)^3} 
 K_{in}(\vec k, \vec p) f(\vec k,t)  f(\vec p,t).
\eea
%
%
 In the limit that 
 $\Gamma_0 + \Gamma(E) \ll k_B T$, we find that 
 the approximate time evolution for $n$ and $T$ is given by
\beq
\partial_t n  =  - \gamma_{in}(t) n, \quad
\partial_t T  =  - \gamma_{in}(t) \Big(\frac{E_{res}}{3 k_B}-\frac{T}{2}\Big)\label{cldynT}
\eeq
with the rate 
\bea\label{gin}
\gamma_{in}(t) & = &
 \frac{1}{\hbar} 2^{3/2} \varpi(n,T)
\frac{\Gamma_0 \Gamma(E_{res})}{\Gamma_0 + \Gamma(E_{res})}
 e^{-E_{res}/(k_B T)},
\eea
 being linear in $n$, and time dependent through $n$, $T$ and $E_{res}$.
 For this approach to be consistent, the thermalization rate 
  during the evolution needs to be larger than the rate $\gamma_{in}(t)$.
 Within the classical theory we find that the thermalization rate 
 is equal to Eq.~\ref{gin} with $\Gamma_0$ replaced by $\Gamma(E)$ in the
 numerator.
 Therefore we have to require $\Gamma(E)/\Gamma_0 \gg 1$, 
 and $\gamma_{in}(t)$ becomes independent of $\Gamma(E)$.

 As an example for the dynamics
 described by Eqs.~(\ref{cldynT}), 
 we consider a process in which the resonance energy
 tracks the temperature at a fixed ratio $\lambda$, i.e.  
 $E_{res}(t) = \lambda k_B T(t)$. 
 We find the solution
\beq
n(t) =  n_0 (1- t/t_{cl})^{\frac{4}{2 \lambda - 7}}, \quad
T(t)  =  T_0 (1- t/t_{cl})^{\frac{2}{3}\frac{2 \lambda - 3}{2 \lambda-7}}
\eeq
 where $n_0$ and $T_0$ are the initial density and temperature,
 and  the classical cooling time $t_{cl}$ is given by
\bea\label{tcl}
1/t_{cl} & = & \frac{2\lambda-7}{4} \gamma_{in}(t=0). 
\eea
The phase space density increases as $\varpi(n,T)= \varpi_0 (1 -  t/t_{cl})^{-1}$,
 where $\varpi_0$ is the initial phase space density. 
 When this approaches one, the system reaches degeneracy and
the classical limit breaks down.
 This occurs shortly before $t_{cl}$, if $\varpi_0\ll 1$.

\begin{figure}
\includegraphics[width=6.4cm]{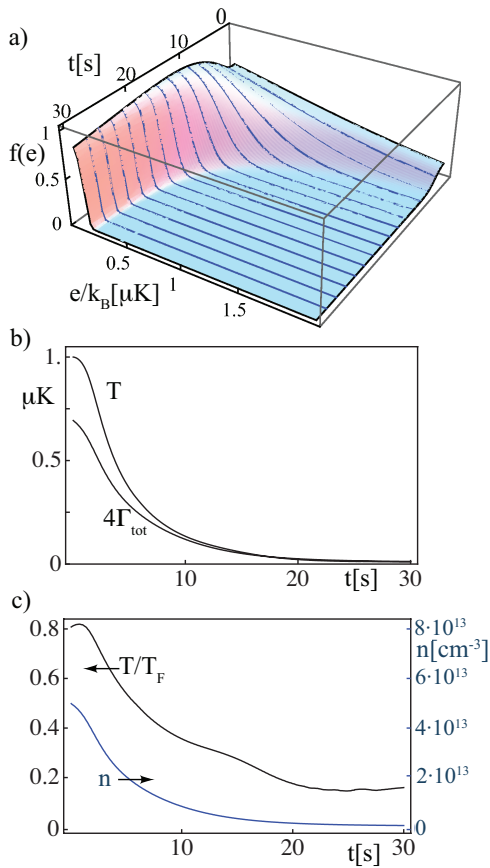}
\caption{\label{cool_single} 
 Collisional cooling of a degenerate Fermi gas with a Feshbach resonance 
  as a function of time for the example described in the text.
 Panel a) shows the distribution $f(e)$ as a function of time.
 Panel b) shows the temperature $T$ and  $4\Gamma_{tot}(2E_F)$.
 Finally, panel c)  shows $T/T_F$  and the density $n$.
}
\end{figure}

For the $^{40}$K example, $\lambda = 11/2$, 
 $T_0 = 1\, \mu$K, and $n_0 = 10^{13}$ cm$^{-3}$, 
 we find $t_{cl} \approx 2$ s.
 This is a realistic time scale for current experimental setups, motivating
  the subsequent, more in-depth analysis.

%

 %
%
 {\it Single resonance cooling.}
 We now consider cooling
  of a gas of spin-polarized fermions 
 in the degenerate regime using a single Feshbach resonance.  
 We use the quantum kinetic theory of Refs.~\cite{luiten, jaksch}. 
  The quantum dynamics is then fully given 
 by the evolution of the momentum state occupation, which satisfies
 a homogeneous quantum Boltzmann equation. 
  We further assume a spherically symmetric momentum
 distribution,
   and the momentum distribution $f$ becomes a function of
 kinetic energy $e = p^2/(2m)$ only. 
 This function $f(e)$ satisfies  
%
%
%
\bea\label{QBE}
\rho(e_1) \partial_t f(e_1) 
 & = & 
- \kappa(e_1) \rho(e_1) f(e_1) - I(e_1)
\eea
where the density of states per unit volume is $\rho(e) = 4 \pi
m^{3/2} \sqrt{2 e}/(2 \pi \hbar)^3$, the loss rate $\kappa$ is given
by
\bea
\kappa(e_1) & = & \frac{1}{2} \int \rho(e_2) d e_2 f(e_2) \int d\cos\theta\, 
   K_{in}(E).
\eea
 We denote $K_{in}(E) \equiv K_{in}(\vec k_1, \vec k_2)$, 
 as the inelastic loss rate coefficient only depends on the
 relative energy  $E = e_1/2 + e_2/2 - \sqrt{e_1 e_2} \cos\theta$, and
 $\theta$ is the angle between $\vec k_1$ and $\vec k_2$.
 
%
%
%
%
The collision integral is given by
\bea\label{collint}
I(e_1) & = &\int d e_2 de_3 W(e_1, e_2, e_3, e_4)\\
& &  \left\{f(e_1) f(e_2) (1 - f(e_3)) (1 - f(e_4))\right.\nonumber\\ 
& & \left. -  (1 - f(e_1)) (1 - f(e_2)) f(e_3) f(e_4)\right\} \nonumber
\eea
 where $e_4 = e_1+e_2 - e_3$, and 
%
%
%
%
%
%
 the collision kernel $W$ is
\bea\label{W}
W(e_1, e_2, e_3, e_4)
& = & \frac{32 \pi^2 m^2}{(2\pi\hbar)^6}
\int_{P_{min}}^{P_{max}} dP\, 
 \sigma(E).
\eea
 The elastic cross-section
  $\sigma(E)=K_{el}(E)/v_r(E)$ and 
  we use $K_{el}(E)\equiv K_{el}(\vec k_1, \vec k_2)$ as it only depends
 on the relative energy $E=e_1+e_2-P^2/(4m)$, where $\vec P$ is
 the total momentum in the collision.
 The integration bounds are given by
%
%
%
%
$P_{min}  =  \max(|p_1 - p_2|, |p_3-p_4|)$ and
 $P_{max}  =  \min(p_1 + p_2, p_3+ p_4)$,
%
%
 where $p_i = \sqrt{2 m e_i}$.
%
%
%
%
%
%
%

%
 We can estimate  the thermalization rate of the system in
 the quantum degenerate regime, based on the Boltzmann equation (\ref{QBE}).
 Following the procedure outlined in [\onlinecite{smithjensen}], 
 we linearize the Boltzmann equation around a
 Fermi distribution $f_0(e)$ with Fermi energy $E_F$ and temperature
 $T$, that is,    
 $f(e) \rightarrow f_0(e) + f_0(e)(1-f_0(e)) \psi(e)$, where
 $\psi(e)$ is a small deviation and the functional form ensures
 that the fluctuations are localized around the Fermi energy.
 Then we find that  the thermalization rate for states close to the
 Fermi energy is given by 
\bea
\frac{1}{\tau_{th}} & \sim & \frac{(k_B T)^2}{\rho(E_F)} W(E_F, E_F, E_F, E_F),
\eea
 reflecting that the only contributions to the collision integral
 of Eq.~\ref{collint} are processes close to the Fermi energy.

The value of $W(E_F, E_F,E_F,E_F)$ 
 can be estimated by realizing that the integral in Eq. \ref{W}
 runs from zero to $2 k_F$, and therefore the relative
 energy from zero to $2 E_F$. As the resonance energy in our cooling scheme
 will be larger than $2 E_F$ by an amount
 of the order of $\Gamma(2 E_F)$, we find
 that $W \propto \sqrt{\Gamma(2 E_F)/E_F}$ and 
   the thermalization
 rate is 
%
%
\bea\label{tauth1}
\frac{1}{\tau_{th}} & \sim & 
 (k_B T)^2 A^{1/2} E_F^{-3/4}/\hbar.
\eea
%
%
%
%
%
%
%
%
%
%
The quadratic temperature dependence is typical for a Fermi gas in
 the degenerate limit.
 For the losses we similarly 
 find  a time scale $1/\tau_l \sim \Gamma_0 \Gamma(2 E_F)/(\hbar E_F)$.

 We now solve the quantum Boltzmann equation numerically to 
 study our cooling process,
 starting from a Fermi distribution for the atoms.
  The initial resonance energy $E_{res}$
  is set well above twice the Fermi energy. 
 We then gradually lower $E_{res}$ to eliminate 
 atoms with large kinetic energy.

The final $E_{res}$, and thus $E_F$,  
 and the time scales of the cooling process can be estimated 
 from our expectation that the smallest $T/T_F$ is of the order
 $\Gamma_{tot}/E_F$. Minimizing this with respect to $E_F$ gives 
 $E_F = (2\Gamma_0/A)^{2/3}$.
 The time scales $\tau_{th}$ and $\tau_l$ will be largest
 at that final value. They are approximately 
 $\tau_{th} \sim \hbar/(A \Gamma_0^{3/2})$ and 
$\tau_l \sim \hbar/(A^{2/3} \Gamma_0^{4/3})$.

   Figure \ref{cool_single} shows an example of cooling for 
 the $^{40}$K resonance described before.
%
%
 The atomic ensemble has an initial temperature of $1.0\, \mu$K and a
 chemical potential of $\mu(0)/k_B = 0.5\, \mu$K, corresponding to an
 initial density of $n\approx 5\cdot 10^{13}$ cm$^{-3}$.  We choose
 $E_{res}(0) = E_{res, i} = 8\, \mu$K, well above $2 E_F \approx 2.5\,
 \mu$K and $T$.  We choose $E_{res, f} = 0.2 \, \mu$K as the final
 value for the resonance energy, based on the above estimate for the
 optimal $T/T_F$.  We use an exponential time dependence $E_{res}(t) =
 (E_{res, i} - E_{res, f}) \exp(- t/t_0) + E_{res, f}$, with a
 timescale $t_0 = 5$ s, that is larger than the estimates for
 $\tau_{th}$ and $\tau_l$.  In Fig.~\ref{cool_single}a we show the
 distribution $f(e)$ as a function of energy and time.  It becomes visibly
  colder in the process.  Throughout the simulation $f(e)$ is fairly
 close to thermal, as expected for $\Gamma(2 E_F)/\Gamma_0\gg 1$.
 Consequently, one can fit $f(e)$ to a Fermi distribution, and assign
 a temperature and a chemical potential at any point in time.
 Figure~\ref{cool_single}b shows the fitted temperature as a function
 of time. It gradually approaches the expected temperature
 $4 \Gamma_{tot}(2 E_F)/k_B$.  The fastest cooling rate or slope,
  around $t\approx 3$ s, is
 consistent with the rate expected in the classical limit, Eq. \ref{tcl}.  The
 cooling is slower for lower temperatures consistent with 
 Eq.~\ref{tauth1}.  The temperature has decreased by two orders of
 magnitude in the process, from $1 \,\mu$K to $10$ nK.  In
 Fig.~\ref{cool_single}c we show $T/T_F$ and $n$.  $T/T_F$ is reduced
 by a factor of $6$, from $0.8$ to $0.15$, while simultaneously the
 density has decreased from $5\times10^{13}$ to $0.7\times 10^{12}$
 cm$^{-3}$.

\begin{figure}
\includegraphics[width=6.0cm]{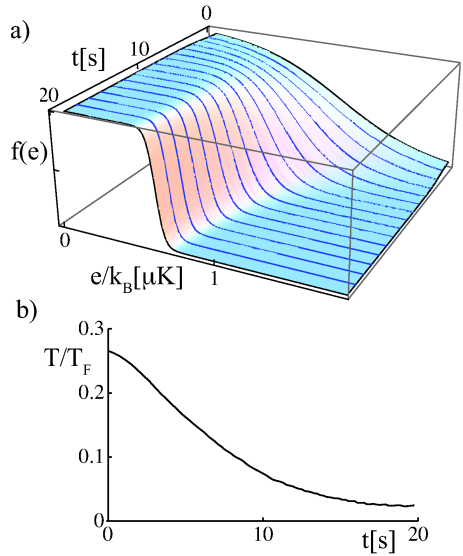}
\caption{\label{cool2restime} 
 Collisional cooling with an optical resonance
 as a ``knife'' and a magnetic resonance for thermalization.
 We use the mass of $^{40}$K, $\Gamma^{opt}(E)/k_B =40\cdot 10^{-6}$ nK$[E/(1$ nK$)]^{3/2}$, 
  and $\Gamma_0^{opt}/k_B =1.8$ nK. 
 For the time dependence of the resonance energy
 we use $E_{res,i}^{opt}=4.1\,\mu$K, $E_{res,f}^{opt}=1.2\,\mu$K, 
  and  $t_0 = 6$ s.
 a) $f(e,t)$.
 b) $T/T_F$.
}
\end{figure}

We have performed multiple simulations with various initial densities,
 temperatures, resonant energies and $t_0$. 
 We observe the same qualitative behavior, except for very small values 
 of $t_0$, where we lose atoms too quickly and the system does not equilibrate.
  Most importantly, we find that $k_B T$ approaches $4 \Gamma_{tot}$.
 By optimizing the functional form of $E_{res}(t)$, this temperature
 could be achieved in a shorter time. However, we do not expect to
 reach significantly lower temperatures.

{\it Cooling with two resonances.}  The elastic and inelastic
scattering rate coefficients are governed by $\Gamma(E)$, which leads
to contradictory requirements.  To overcome this limitation, we
propose the use of two $p$-wave resonances:  a narrow optical or rf induced
 one, which acts
as the 'knife', and a broad loss-less magnetic one which 
thermalizes.  We locate the magnetic resonance  such that the
elastic rate coefficient is unitarity limited at $K_{el} = v_r \pi
\hbar^2/k_r^2$. 
 To ensure a loss-less magnetic resonance the atoms must be in the
 lowest hyperfine state and the
field driving the
 narrow transition must not couple to the Feshbach
molecular state \cite{kostrun}.
 On the other hand, 
 the loss from and the  width of  
 the optical (rf) resonance can be controlled by the 
 laser intensity (rf field). This 
 creates a very narrow ``knife'' with a negligible contribution
 to the elastic rate coefficient.
 Its resonance
  location $E_{res}^{opt}$ is lowered in time.

We have repeated the analysis of the previous section.
The thermalization rate is driven by the 
 magnetic resonance and given by $1/\tau_{th} \sim (k_B
 T)^2/(\hbar E_F)$.
  Hence,   
 unlike for the single resonance case,
 thermalization does not require $\Gamma^{opt}(E)/\Gamma_0^{opt}\gg
 1$.  In fact we find that $\Gamma^{opt}(2E_F)\approx\Gamma_0^{opt}$ leads to the 
 fastest cooling for a given $\Gamma^{opt}_{tot}$. 
 Figure~\ref{cool2restime} shows an example of the cooling process. 
 The initial density is $4\times 10^{13}$ cm$^{-3}$, the initial temperature
 is $0.3\, \mu$K. The system is cooled down to a final density 
 $1.9\times 10^{13}$ cm$^{-3}$ and a final temperature $0.017\, \mu$K.
 This  demonstrates that
 $T/T_F \approx 2.5\times10^{-2}$ can be achieved with this cooling process.

In conclusion, we have proposed a new cooling mechanism which uses
 the energy selectivity of a Feshbach resonance.
 We first discussed the limit of a classical thermal gas, 
 before we turned to 
 a quantum kinetic simulation of a degenerate Fermi gas.
 The case of a single resonance 
shows cooling to the regime around $T/T_F\approx 0.1$ for
 $^{40}$K, 
 for an appropriately chosen resonance.
 We then improve on this setup by using 
 one narrow resonance for the loss process and a broad magnetic
 resonance for thermalization. 
 This setup can create temperature regimes competitive  with
 current technology.

This work was supported by NSF under Physics Frontier Grant PHY-0822671.
 L.M. acknowledges support from an NRC/NIST fellowship.

\def\etal{\textit{et al.}}

\end{document}